\documentclass[aps,prb,twocolumn,floatfix,superscriptaddress,showpacs,amsmath,amssymb]{revtex4-1}
\usepackage{bm}
\usepackage{amssymb}
\usepackage{colordvi}
\usepackage{graphicx}
\usepackage{color}
\usepackage{hyperref}

\newcommand{\ov}[1]{\overline{{#1}}}
\newcommand{\be}{\begin{equation}}
\newcommand{\ee}{\end{equation}}
\newcommand{\bea}{\begin{eqnarray}}
\newcommand{\eea}{\end{eqnarray}}

\newcommand{\vep}{\varepsilon}

\newcommand{\ome}{\omega}
\newcommand{\Ome}{\Omega}

\def\nn{\nonumber}

\begin{document}

\title{Phonon-thermoelectric transistors and rectifiers}

\author{Jian-Hua Jiang}
\affiliation{Department of Physics, Soochow University, 1 Shizi Street,
 Suzhou 215006, China}
\author{Manas Kulkarni}
\affiliation{Department of Physics, New York City College of Technology, The City University of New York, Brooklyn, NY 11201, USA}
\author{Dvira Segal}
\affiliation{Chemical Physics Theory Group,
Department of Chemistry, University of Toronto, 80 Saint George
Street, Toronto, ON, M5S 3H6, Canada}
\author{Yoseph Imry}
\affiliation{Department of Condensed Matter Physics, Weizmann Institute of Science, Rehovot 76100, Israel}

\date{\today}

\begin{abstract}
We describe {\em nonlinear} phonon-thermoelectric devices where charge
current and electronic and phononic heat currents are coupled, driven
by voltage and temperature biases, when phonon-assisted inelastic
processes dominate the transport. Our thermoelectric transistors and
rectifiers can be realized in a gate-tunable double quantum-dot system
embedded in a nanowire which is realizable within current technology.
The inelastic electron-phonon scattering processes are found to induce pronounced charge,
heat, and cross rectification effects, as well as a thermal transistor effect that, remarkably,
can appear in the present model even in the linear-response regime
without relying on the onset of
negative differential thermal conductance.
%
\end{abstract}


\maketitle

\section{Introduction}
Onsager's formulation of irreversible thermodynamics and his famous
reciprocal relations (1931)\cite{Onsager}, based on
the time reversibility of microscopic dynamics, tie different currents
to driving forces (affinities) and provide strict bounds on the
efficiency of energy transducers \cite{PhysRevE.90.042126}.
Among such transport phenomena the thermoelectric effect
which describes the coupling of electronic charge and heat currents
nowadays attracts significant attention experimentally, computationally, and fundamentally, %
as new-unique thermoelectric materials and devices are pursued
with a promise for an improved efficiency, 
see e.g. Refs. \onlinecite{Dresselhaus,Majumdar10,Segalman,Hochbaum,Dubi,ShakouriRev,JordanRev}.
%
The thermoelectric effect has been traditionally characterized by linear response quantities,
Seebeck and Peltier coefficients.
{\it Nonlinear} thermoelectric phenomena constitute a new area of 
research, anticipated to enhance thermoelectric response \cite{shakouri4,
shakouri7,nonlinear1,nonlinear2}. 
An elastic scattering theory of nonlinear thermoelectric transport has been recently developed
by considering self-consistent screening potentials \cite{nonlinear1,sanchez1,sanchez2,sanchez3,sanchez4,Linke,Whitney1,Whitney2}.
Other studies had considered nonlinear thermoelectric transport with explicit electron-electron \cite{Kuo,Dare},
electron-phonon \cite{nonlinear2} and electron-photon \cite{Lehur15} interactions.
However, an investigation of nonlinear inelastic
``phonon-thermoelectric'' systems, where charge current and electronic and
phononic heat currents are coupled, and non-linear thermoelectric
device operations (such as rectifiers and transistors) is still
missing.

\begin{figure}[]
\includegraphics[scale=.75]{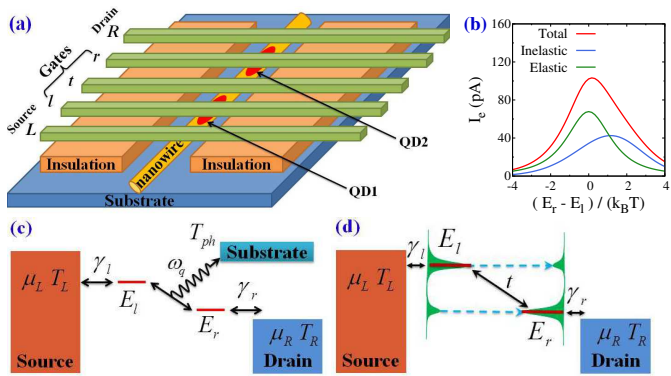}
\caption{(Color online) (a) Scheme of a DQD that can serve as a
phonon-thermoelectric diode and transistor. The QDs are embedded
in the nanowire and are controlled by gate voltages: $l$ and $r$ control
the local potentials, $t$ tunes the tunneling between the
QDs. The two electrodes, $L$ and $R$, apply voltage and temperature
biases across the QDs. The insulation layer suppresses the thermal contact
between the metal electrodes and the substrate, which provides
thermal energy to phonons. 
(b) Elastic and inelastic contributions to the charge current
as a function of level detuning; 
microscopic processes are illustrated in panels (c) and (d).
(c) Inelastic thermoelectric transport assisted by a phonon of frequency $\ome_q$.
(d) Elastic thermoelectric transport. The shaded
green area represents the broadening of the left and right QDs including the
hybridization between them. The dashed arrows display the two main
tunneling paths, of different energies. In (b) we used $t=15~\mu$eV,
$\xi_0=1~\mu$eV, $\ome_0=100~\mu$eV, $\gamma_{\ell}=\gamma_r=5~\mu$eV,
$k_BT=20~\mu$eV (for $T_L=T_R=T_{ph}=T$), $E_\ell=0$, and $\mu_L=-\mu_R=40~\mu$eV.}
\label{fig:fig1}
\end{figure}

{By ``phonon-thermoelectric'' systems, we refer to a setup first
considered in Ref.~\onlinecite{ora} (see also, Refs.~\onlinecite{prb1,njp,jap-j}), where electrical and thermal heat
currents from source to drain were induced and manipulated by a third, independent, phonon reservoir.
This terminal is
characterized by its temperature, which is possibly distinct from the temperatures of the source and drain.
The physical mechanisms involving the phonon baths are
inelastic-phonon assisted electron hopping from the source to the drain.
In the setup shown in Fig. \ref{fig:fig1}(a),
the substrate corresponds to the third terminal, and it determines the temperature of phonons.
In the linear response regime, a
hot phonon bath can pump electrical current, to drive electrons against the
electrochemical potential gradient. This effect was termed as the
``three-terminal thermoelectric effect'', 
and it is similar to the photo-electric effect in photovoltaic systems where
high-temperature photons pump electric currents. In fact, this
analogy has been utilized for proposing useful thermoelectric
devices\cite{njp}. Here, we extend this mechanism to the
nonlinear response regime, to realize nontrivial thermoelectric
functions and devices.}


Nonlinear {\it electrical} elements, diodes, amplifiers, and transistors,
are key components in electronics, at the heart of modern technology.
More recently, significant efforts have been devoted to
the exploration of analogous {\it phononic} elements at the nanoscale \cite{bli}.
These endeavors, nonlinear nanoelectronic and nanophononic, were often
separately pursued. However, it is to be noted that in many cases
the interaction of electrons with the atomic degrees of freedom
(vibrations, phonons) cannot be ignored. While electron-phonon dissipation effects often impede
device operation, here we exploit them to achieve compound nonlinear
functionalities.
{We demonstrate these functionalities in a
three-terminal geometry (the simplest situation). The effects described here can
be generalized to other situations in which electronic currents and
phononic currents are spatially separated.}

%

In this work, we provide a unified description of coupled and
nonlinear electrical and phononic transport in the three-terminal
geometry when inelastic electron-phonon scatterings play a key role.
The nonlinear, inelastic electron transport in our geometry (see
Fig. \ref{fig:fig1}) provides charge and thermal rectification,
cross-rectification effects (e.g., thermal rectification induced by voltage),
as well as the realization of the thermal transistor effect and a phonon-thermoelectric cross
effect (e.g., controlling source-drain $I_e$-$V$ characteristics by
modulating the temperature of a third terminal). Remarkably, we find
that in the {\em linear response} regime {\em inelasticity} can lead
to a thermal transistor effect. This is in contrast to common wisdom
\cite{liwang,bli} that thermal transistor effect must comply with
negative differential thermal conductance. The essential ingredient here is that
inelastic electron-phonon scattering processes simultaneously
involve three different reservoirs. This mechanism also works for
inelastic electron transport assisted by coulomb drag effects
\cite{coulomb1,coulomb2}, magnons \cite{magnon}, photons
\cite{photon,microwave}, and plasmons \cite{plasmon}, thus our results
are of interest beyond the specific model considered in this work.



\section{Model system}
We consider a double quantum-dot (DQD) embedded
in a nanowire in contact with two metals and a phonon substrate (``$ph$"), see Fig. \ref{fig:fig1}.
The quantum dots (QDs) are defined by voltage gates
(labeled by $l$ and $r$ in Fig. \ref{fig:fig1}a) with tunable electronic energy levels
$E_l$ and $E_r$. $t$ is a hopping element between the QDs
and $\gamma_{l,r}$ are the hybridization energies
of the dots to the source
and drain electrodes, labeled by $L$ and $R$, respectively.
Charge current,  electronic heat current, and phononic heat current are induced by
applying a voltage bias between the terminals $L$ and $R$ and a
temperature difference between the three terminals.
This setup (and related models) have been explored  in e.g., Refs.
\onlinecite{JKoch,ds,Linano12,manas,thermalgating,TKoch,Felix,Linano15,ds15,ora,prb1,njp}.
However, previous studies had focused on the linear response regime, while here we uncover
coupled {\em nonlinear} phenomena. The system is described by the Hamiltonian,
\be
\hat H = \hat H_{DQD} + \hat H_{e-ph} + \hat H_{lead} + \hat H_{tunel} + \hat H_{ph}
\ee
with
\begin{subequations}
\begin{align}
& \hat H_{DQD} = \sum_{i=\ell,r} E_i \hat c_i^\dagger \hat c_i + ( t \hat c_\ell^\dagger \hat c_r +
{\rm H.c.}) , \\
& \hat H_{e-ph} = \sum_{ q} U_{ q} \hat c_\ell^\dagger \hat c_r
(\hat a_{ q} + \hat a^\dagger_{- q}) + {\rm H.c.} ,\\
& \hat H_{lead} = \sum_{j=L, R}\sum_{k} \vep_{j,k} \hat c_{j,k}^\dagger \hat c_{j,k} ,\\
& \hat H_{tunel} = \sum_k V_{L, k} \hat c_\ell^\dagger \hat c_{L, k} + \sum_k V_{R, k}
\hat c_r^\dagger \hat c_{R, k} +  {\rm H.c.} , \\
& \hat H_{ph} = \sum_{ q} \ome_{q}(\hat a_{q}^\dagger \hat a_{q} +\frac{1}{2}) .
\end{align}
\end{subequations}
Here $\hat c_i^\dagger$ ($i=\ell, r$) creates an electron in the $i$-th
QD with an energy $E_{i}$.
The $\ell$ ($r$) QD is located next to the left (right) lead.
{The tunneling elements from the $\ell$ QD to the right lead and
  that from the $r$ QD to the left lead are assumed negligible.}
Electron-phonon interactions (matrix element $U_{q}$) allow inelastic electron transport through the system;
$\hat a_{q}^\dagger$ creates a phonon with a wavevector
$q$ and frequency $\ome_{q}$.
We adopt the conventions that
$\hbar\equiv1$ and $k_B\equiv 1$ throughout the paper.
{To maintain a finite temperature difference between the electrodes and
the substrate, we suggest to thermally isolate them with a layer of
thermal and electrical insulation [see Fig. \ref{fig:fig1}(a)].}
The phonons involved in the inelastic
transport through the DQD system may be confined phonons in the
quantum wire, or bulk phonons in the substrate. We
assume the thermal contact between the quantum wire and the substrate
to be good so that these two types of phonons acquire the same
temperature $T_{ph}$. The phonon temperature can be controlled by the
substrate, while the temperature of the electrodes ($T_L$ and $T_R$)
can be individually controlled by heating them  with, e.g.,
AC electric fields \cite{giazotto,Dubi}.
{Although Coulomb
  interaction is certainly important in electron transport effects through QDs, to
  simplify the problem and to uncover the essential physics, we assume
  that the intra-dot Coulomb interaction energy is much
  larger than other relevant energy scales in our problem, as well as that the
  inter-dot Coulomb interaction is small and negligible. Under these
  assumptions, we can ignore the Coulomb interaction altogether. This
  assumption limits our discussions to the regime where there is no
  more than one electron in each QD. Nevertheless, the
  essential element for the phenomena described in this work is energy
  exchange between electrons and phonons, rather than electron-electron energy exchange process.}

\begin{figure}[h]
\includegraphics[scale=0.92]{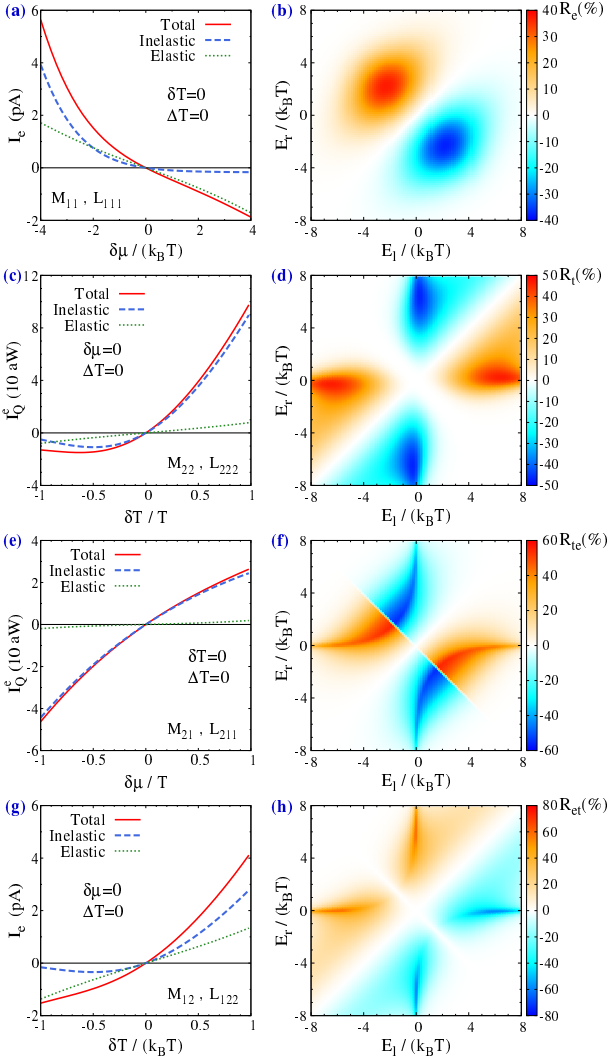}
\caption{(Color online) Charge, heat,  and cross rectification effects.
(a) Charge current $I_e$ for $E_\ell=-E_r=-2k_BT$ and (b)
optimization of charge rectification $R_e$
with QDs energies $E_\ell$  and $E_r$ at  $\delta\mu=20~\mu$eV.
(c) Electronic heat current $I_Q^e$ ("aW" denoting atto-Watts) for
$E_\ell=-5k_BT$ and $E_r=-k_BT$, and (d) optimization of the associated rectification strength $R_t$, 
by tuning $E_\ell$ and $E_r$ using $\delta T=0.5T$.
(e) Electronic heat current as a function of applied voltage for $E_\ell=-k_BT$ and
$E_r=2k_BT$,
and (f) optimization of the relevant rectification $R_{te}$ at the bias $\delta\mu=20~\mu$eV.
(g) Charge current against the (metals) temperature difference for
  $E_\ell=-5k_BT$ and $E_r=-k_BT$, and (h) optimization of the
 rectification strength $R_{et}$
by controlling the DQDs at $\delta T=0.5T$.
We used $t=15~\mu$eV, $\xi_0=1~\mu$eV, $\ome_0=100~\mu$eV,
  $\gamma_{\ell}=\gamma_r=5~\mu$eV, $k_BT=20~\mu$eV, and
  $T_{ph}=T_0$ for all figures.
}
\label{fig:fig2}
\end{figure}

\section{Currents in and beyond linear-response}
 Non-equilibrium
steady state quantities of interest are the electric current $I_e$,
the electronic heat current traversing from the left lead to the right
lead $I_Q^{e}\equiv \frac{1}{2}(- \dot Q_L+\dot Q_R)$, and the phonon heat
current $I_Q^{ph}\equiv - \dot Q_{ph}$, with $\dot Q_{i}$ ($i=L,R,ph$) denoting the
heat current flowing into the $i$th reservoir.
Both elastic and inelastic processes contribute to transport in the system [see
 panels (b), (c) and (d) in Fig.\ref{fig:fig1}].

The inelastic (``{\it inel}") contribution to the currents is calculated from
 the Fermi Golden-Rule treatment of Ref. \onlinecite{prb1},
setting for convenience $\mu\equiv \frac{1}{2}(\mu_L+\mu_R)=0$,
\begin{align}
I_e|_{inel} = eI_N, \quad I_{Q}^e|_{inel} = \ov{E} I_N , \quad
I_{Q}^{ph}|_{inel} = \Ome I_N
\label{in-I}.
\end{align}
Here $e<0$ is the charge of an electron, $\ov{E}\equiv\frac{E_{\ell} +
  E_r}{2}$, and $\Ome \equiv E_r-E_{\ell}$. In the above equation,
$I_N = \Gamma_{\ell\to r} - \Gamma_{r\to \ell}$ with $\Gamma_{\ell\to
  r} \equiv \gamma_{e-ph} f_{\ell} ( 1 - f_r) N_p^-$ and $\Gamma_{r\to
  \ell} \equiv \gamma_{e-ph} f_r (1-f_{\ell}) N_p^+$. Here
$N_p^{\pm}=N_B+\frac{1}{2}\pm\frac{1}{2}{\rm sgn}(\Ome)$
with the Bose-Einstein distribution for phonons $N_B\equiv
[\exp(\frac{|\Ome|}{T_{ph}})-1]^{-1}$. {For the commonly used zinc
  blend semiconductors such as GaAs and InP}, the electron-phonon
interaction energy $U_q$ determines the transition
rate\cite{dqd1,dqd2,weber,manas}  $\gamma_{e-ph}= \xi_0
\left(\frac{|\Ome|}{\ome_{0}}\right)^{n}
\exp\left[-\left(\frac{\Ome}{\ome_{0}}\right)^2\right]$. Here $\xi_0$
stands for the electron-phonon scattering strength, $n$ provides the
power-law dependence on phonon energy $\Ome$ with a characteristic
energy $\ome_0$.
{
Appendix A provides details on the derivation of this expression
  from the microscopic theory of electron-phonon
  interactions in GaAs QDs. It is found that in this system the piezoelectric
  mechanism dominates the electron-phonon interaction.}
In this
situation, we adopt the parameters $\xi_0=1~\mu$eV and $n=1$
\cite{weber,manas}. The Fermi Golden Rule method is valid when $k_BT
\gg \gamma_{\ell},\gamma_{r}\gg | \Gamma_{\ell\to r} - \Gamma_{r\to
  \ell}|$ \cite{prb1}.
{These conditions can be maintained by tuning the coupling
between the QDs and the leads}. In this regime, {the steady state
distributions on the two QDs can be approximated by those of the
nearby leads \cite{prb1}, i.e., $f_{\ell}\simeq f_L(E_\ell) =
[\exp(\frac{E_{\ell}-\mu_L}{T_L})+1]^{-1}$ and $f_r\simeq f_R(E_r) =
[\exp(\frac{E_r-\mu_R}{T_R})+1]^{-1}$, where $f_L$ and $f_R$ are the
Fermi distribution functions for the left and right electronic
reservoirs, respectively.
If the Fermi Golden Rule assumptions (specifically, the second inequality) are not
satisfied, the inelastic current can be calculated more generally via a rate equation
technique [see Appendix B]. The Fermi Golden Rule method is
exploited below for the analytic study of transport coefficients. Numerical
results in this work are calculated directly from the rate equation method
(unless specified otherwise).}

The contribution of elastic (``{\it el}") processes to the currents is given by Landauer's formula
\begin{subequations}
\begin{align}
I_e|_{el} &= e \int \frac{d\vep}{2\pi} {\mathcal T}(\vep)[f_L(\vep) - f_R(\vep)] ,\\
I_{Q}^e|_{el} &= \int \frac{d\vep}{2\pi}\vep {\mathcal T}(\vep)[f_L(\vep) - f_R(\vep)]
,\quad I_{Q}^{ph}|_{el} = 0.
\end{align}
\label{el-I}
\end{subequations}
{Note that elastic processes do {\em not} contribute to the
  heat current from the phonon terminal $I_{Q}^{ph}$.}
The energy-dependent transmission function is obtained from the Caroli formula,
${\mathcal T}(\vep) = {\rm
  tr}[\hat{G}^r(\vep)\hat{\Gamma}_L\hat{G}^a(\vep)\hat{\Gamma}_R] $,
where $\hat{G}^r(\vep) = \hat{G}^{a \dagger}(\vep) =
\left[ \begin{array}{cccc} \vep - E_{\ell} + i\gamma_{\ell}/2 & -t \\
    -t & \vep - E_r + i\gamma_{r}/2
  \end{array}\right]^{-1}$, $\hat{\Gamma}_L = \left[ \begin{array}{cccc}
    \gamma_{\ell} & 0 \\
    0 & 0
  \end{array}\right]$, and $\hat{\Gamma}_R =
\left[ \begin{array}{cccc}
    0 & 0 \\
    0 & \gamma_{r}
  \end{array}\right]$,
where we assume constant (energy-independent) tunneling rates $\gamma_\ell$ and $\gamma_r$.
This results in ${\mathcal T}(\vep) =
t^2\gamma_{\ell}\gamma_{r}/|d_G|^2$ with $d_G =
(\vep-E_{\ell} + i\gamma_{\ell}/2)(\vep-E_r + i\gamma_{r}/2) - t^2$.
In Fig. \ref{fig:fig1}(b) we plot the elastic contribution to the charge
current (green) as a function of the DQD  level detuning for an otherwise symmetric junction
$\gamma_l=\gamma_r$. As expected, the current is an even function with respect to detuning.
In contrast, the inelastic
contribution (blue) is {\em asymmetric}
with respect to detuning. This distinction serves as a signature of inelastic
transport in DQD systems in experiments \cite{manas}.
{Contributions
from elastic transport processes are included in our simulations.}

The entropy production rate for the whole system is given by
\be
\dot S_{tot} = \frac{\dot Q_L}{T_L} + \frac{\dot Q_R}{T_R} +
\frac{\dot Q_{ph}}{T_{ph}} = \sum I_i A_i, \label{eq-5}
\ee
where the affinities for $I_e$, $I_Q^e$, and $I_Q^{ph}$ [identified in
Eq. (\ref{eq-5}) by $I_1, I_2, I_3$ separately], are
$A_1 = \frac{\mu_L-\mu_R}{e}(\frac{1}{2T_L}+\frac{1}{2T_R}),\quad
A_{2} = \frac{1}{T_R} - \frac{1}{T_L}, \quad A_{3} = \frac{1}{2T_L}
+\frac{1}{2T_R}- \frac{1}{T_{ph}}$, respectively. By setting the reference
temperature at $T_0\equiv \frac{2T_LT_R}{T_L+T_R}$, we obtain
$A_1=\frac{\delta\mu}{eT_0}$, $A_2=\frac{\delta T}{T_LT_R}$, and
$A_3=\frac{\Delta T}{T_0T_P}$ where $\delta \mu \equiv \mu_L - \mu_R$,
$\delta T \equiv T_L - T_R $, and $\Delta T \equiv T_{ph}-T_0$
\cite{prb1}. Note that the reference temperature is defined according
to $\frac{1}{T_0}\equiv \frac{1}{2}(\frac{1}{T_L}+\frac{1}{T_R})$ [differing
from the average temperature $T\equiv\frac{1}{2}(T_L+T_R)$ in
nonlinear-response regime]\cite{nl-Onsager}.

Expanding Eqs. (\ref{in-I}) and (\ref{el-I}) to capture
nonlinear effects, we find, to the lowest nontrivial order
\begin{align}
I_i = \sum_j M_{ij} A_j + \sum_{jk} L_{ijk} A_j A_k + {\cal O} (A^3) , \label{trans-non}
\end{align}
where $M_{ij}=M^{el}_{ij} + M^{inel}_{ij}$ with $M_{ij}$ denoting
the linear-response coefficients. The Onsager reciprocal relations,
$M_{ij}=M_{ji}$, hold for both elastic and {\em inelastic} transport
processes. The second-order terms $L_{ijk}$
only come up from the {\em inelastic} transport processes.
Elastic coefficients are given by
\begin{subequations}
\begin{align}
& M^{el}_{11} = \frac{e^2}{2\pi} \int d\vep {\mathcal T}(\vep) f^{eq}(\vep) [1-f^{eq}(\vep)],
\\
& M^{el}_{12} = M^{el}_{21} = \frac{e}{2\pi} \int d\vep {\mathcal T}(\vep)
f^{eq}(\vep) [1-f^{eq}(\vep)] \vep, \\
& M^{el}_{22} = \frac{1}{2\pi} \int d\vep {\mathcal T}(\vep)
f^{eq}(\vep) [1-f^{eq}(\vep)] \vep^2 ,\\
& M^{el}_{3j} = M^{el}_{j3} = 0,\quad j=1,2,3.
\end{align}
\end{subequations}
The superscript ${eq}$ denotes the equilibrium distribution.
The linear response inelastic coefficients which satisfy the Onsager
reciprocal relations are
\be
M^{inel}_{ij} = s_i s_j \Gamma^{eq}_{\ell \to r},
\ee
where $\Gamma^{eq}_{\ell \to r}=\gamma_{e-ph}
f_{\ell}^{eq} ( 1 - f_r^{eq}) (N_p^-)^{eq}$ is the
transition rate from QD $\ell$ to QD $r$ at equilibrium, and $s_1
= e, \quad s_2 = \ov{E} , \quad s_3 = \Ome$.

The {\em second-order}
transport coefficients are given by
\be
\label{lijk}
L_{ijk} = \frac{1}{2}s_i g_{jk} \Gamma^{eq}_{\ell \to r},
\ee
where ($g_{jk}=g_{kj}$)
\begin{align}
g_{11} &= e^2(f^{eq}_r - f^{eq}_{\ell}) , \nn \\
g_{12} &= e \ov{E} (f^{eq}_r - f^{eq}_{\ell}) +
\frac{1}{2} e\Ome ( f^{eq}_{\ell} +  f^{eq}_r -1 ), \nn \\
g_{13} &= 0, \quad \quad g_{23} =  0 , \nn \\
g_{22} &= (\ov{E}^2+ \frac{1}{4}\Ome^2) (f^{eq}_r - f^{eq}_{\ell}) + \ov{E} \Ome  (
f^{eq}_{\ell} +  f^{eq}_r -1 )  ,\nn \\
g_{33} &= 2 \Ome |\Ome| (1 + 2 N_B^{eq}) .
\label{gijk}
\end{align}
Note that Onsager reciprocal symmetry does {\em not} guarantee
$L_{ijk}=L_{jik}$ \cite{nl-Onsager}.

{For the described functionalities to be prominent, the
  contribution from the inelastic processes needs to be enhanced while
  the elastic processes must be suppressed. This is realized by strong
  electron-phonon interaction, high temperature\cite{prb1}, and the
  mismatch of QDs energy levels. The latter creates a barrier for
  elastic tunneling and favors inelastic transport. We noticed from
  Fig. \ref{fig:fig1}(b) that there is an optimal alignment of QDs
  energy levels for inelastic processes to be dominant. It is also noted that the
inelastic contribution is nonzero even at $\Ome=0$, which is a special
property of the piezoelectric electron-phonon interaction: The electron-phonon coupling
strength is proportional to $\Ome$, but this linear dependence 
cancels out the divergence in the  phonon number in the limit $\Ome\to 0$,
overall leaving a finite term which contributes to the current. A careful
examination of the $\Ome\to 0$ limit for realistic systems is beyond
the scope of this paper. Our main conclusions do not rely on
the behavior of the system in this special regime.}

The transport coefficients depend on the average energy $\ov{E}$
(representing deviation from particle-hole symmetry) and the detuning $\Ome$
(measuring deviation from mirror symmetry). The symmetry of transport
coefficients and currents under (i) particle-hole transformation:
${\cal T}_{PH}: E\to -E$ (hence $\ov{E}\to -\ov{E}$ and $\Ome\to
-\Ome$) and (ii) mirror symmetry (assuming $\gamma_\ell=\gamma_r$)
${\cal T}_{M}: E_{\ell}\leftrightarrow E_r$ (i.e., $\Ome\to -\Ome$)
are summarized in Table~I. These symmetries rule out certain
nonlinear effects in the presence of mirror symmetry and/or particle-hole
symmetry. For example, when $\ov{E}= 0$ but $\Ome\neq 0$
only terms of the same symmetry under the
particle-hole and mirror transformations remain nonzero.

\begin{table}[htb]
\caption{Symmetries of $g_{ij}$, $L_{ijk}$, $M_{ij}$, and the heat currents.}
\begin{tabular}{llllllllll}\hline 
Terms  & \mbox{} & Particle-hole ${\cal T}_{PH}$ & \mbox{} &
Mirror ${\cal T}_{M}$ \\
\hline
$g_{11}$, $g_{22}$, $g_{33}$ & \mbox{} & Odd & \mbox{} & Odd \\
$g_{12}$ & \mbox{} & Even & \mbox{} & Odd \\
$L_{111}$, $L_{122}$, $L_{133}$, $L_{212}$ & \mbox{} & Odd & \mbox{} & Odd \\
$L_{112}$, $L_{211}$, $L_{222}$, $L_{233}$  & \mbox{} & Even & \mbox{} & Odd \\
$L_{311}$, $L_{322}$, $L_{333}$ & \mbox{} & Even & \mbox{} & Even \\
$L_{312}$, $I_Q^{ph}$, $M_{12}$ & \mbox{} & Odd & \mbox{} & Even \\
$I_Q^e$, $M_{13}$ & \mbox{} & Odd & \mbox{} & Odd \\
$M_{23}$ & \mbox{} & Even & \mbox{} & Odd \\
$M_{11}$, $M_{22}$, $M_{33}$ & \mbox{} & Even &
\mbox{} & Even \\
\hline 
\end{tabular}
\label{table1}
\end{table}

\begin{table}[htb]
\caption{Functionality of second-order coefficients}
\begin{tabular}{llllllllll}\hline 
Terms ($L_{ijk}$) & \mbox{} & Diode or Transistor effect \\
\hline
$L_{111}$ &  \mbox{} & charge rectification \\
$L_{222}$, $L_{333}$ &  \mbox{} & electronic and phononic heat rectification \\
$L_{233}$, $L_{322}$ & \mbox{} & off-diagonal heat rectification \\ 
$L_{122}$, $L_{133}$ & \mbox{} & charge rectification by
$\delta T$ or $\Delta T$\\
$L_{211}$, $L_{311}$ &  \mbox{} & heat rectification by voltage\\
$L_{212}$, $L_{112}$ &  \mbox{} &  other nonlinear thermoelectric \\
$L_{321}$ & \mbox{} & phonon-thermoelectric transistor \\ 
\hline 
\end{tabular}
\label{table2}
\end{table}

\section{Rectification (diode) effect}
Coupled thermal and electrical transport allows unconventional
rectification, for example, charge rectification induced by a
temperature difference, besides showing the standard charge and heat
rectification effects. The magnitude of the rectification effect is
defined by $R_e=\frac{I_e(V)+I_e(-V )}{|I_e(V )|+|I_e(-V )|}$ for
charge rectification, $R_t=\frac{I_Q^e(\delta T)+I_Q^e(-\delta
  T)}{|I_Q^e(\delta T)|+|I_Q^e(-\delta T)|}$ for (electronic) heat
rectification, $R_{et}=\frac{I_e(\delta T)+I_e(-\delta T)}{|I_e(\delta
T)|+|I_e(-\delta T)|}$ for charge rectification induced by the temperature
difference $\delta T$, and
$R_{te}=\frac{I_Q^e(V)+I_Q^e(-V)}{|I_Q^e(V)|+|I_Q^e(-V)|}$ for heat
rectification induced by voltage bias.
We shall refer to $R_{et}$ and $R_{te}$ as ``thermoelectric
rectifications". The dependence of these quantities on the DQDs
energies is displayed in Fig. \ref{fig:fig2}, demonstrating
significant rectification effects.

From these figures, one finds that thermal diode $R_t$ and thermoelectric diode
$R_{te}$ cannot be realized when the system is invariant under ${\cal
  T}_{PH}\otimes {\cal T}_{M}$ (i.e., vanishing $\ov{E}$).
In Table II we classify second-order transport coefficients
as rectifiers and transistors.
{We note that the rectification and
transistor behavior as displayed in  Fig.\ref{fig:fig2} [obtained from a direct calculation
using Eqs. (\ref{in-I})-(\ref{el-I})],
acquire the same symmetry as the corresponding second-order transport
coefficients listed in Table~\ref{table2}. Specifically, $R_e$ and $R_{et}$
(relating to $L_{111}$ and $L_{122}$, respectively)
are odd under both ${\cal T}_{PH}$ and ${\cal T}_M$, whereas $R_t$ and
$R_{te}$ (relating to $L_{222}$ and $L_{211}$, respectively) are
even under ${\cal T}_{PH}$, but odd under ${\cal T}_M$. In addition,
it is demonstrated in Fig.\ref{fig:fig2} that one can tune  $E_\ell$
and $E_r$ for reaching optimal electrical, thermal, and cross rectifications.
}

\begin{figure}[htbp]
\includegraphics[scale=1.]{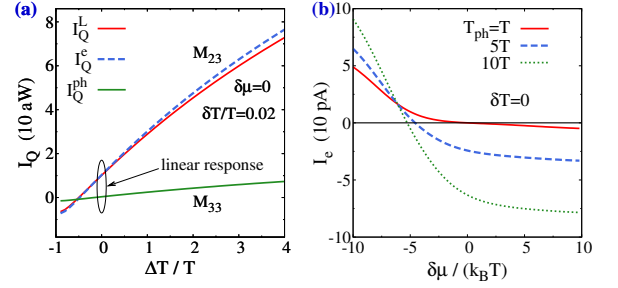}
\caption{(Color online) (a) Linear thermal transistor.
We plot the electronic heat currents  $I_Q^L$  and $I_Q^e$
(the former describes the current leaving the $L$ terminal)
and the phonon heat current
$I_Q^{ph}$ ("aW" denoting atto-Watts),  against the temperature of the phonon bath. The two electronic heat
currents vary significantly whereas the phonon current changes
very little with $T_{ph}$.
$E_\ell=4.5k_BT$, $E_r=5k_BT$, $\delta T=0.02 T$, $k_BT=20~\mu$eV, and
$\delta\mu=0$. 
(b) Phonon-transistor. The charge current $I_e$,  plotted
as a function of $\delta\mu$, is
largely controlled by the temperature of the phonon bath.
$E_\ell=-4k_BT$, $E_r=2k_BT$, and $\delta T = 0$.
Other parameters are the same as in Fig.\ref{fig:fig2}.}
\label{fig:fig3}
\end{figure}

\section{Transistors}
One of our central results
is that a thermal transistor effect can develop in the {\em linear-response regime}.
Specifically, the heat current amplification factor is given by
\be
\alpha \equiv \left |\frac{\partial_{T_{ph}} I_Q^L}{\partial_{T_{ph}}
    I_Q^{ph}} \right| =\frac{|M_{23}-\frac{1}{2}M_{33}|}{M_{33}} = \left | \frac{E_\ell}{\Ome} \right |  \label{aph},
\ee
where $I_{Q}^L=I_Q^e-I_Q^{ph}/2$ is the heat
current flowing out of the $L$ electrode (since $\mu_\ell=\mu_r$).
Eq. (\ref{aph}) follows directly from  (i) $I_Q^L=E_\ell I_N + ``{\sl
  elastic \ contributions}"$, $I_Q^{ph}=\Ome I_N$ [according to
Eq.~(\ref{in-I})], and (ii) the elastic contribution to $I_Q^L$ does
not depend on the phonon temperature $T_{ph}$. Hence
\begin{align}
& \partial_{T_{ph}}I_Q^L=E_\ell \partial_{T_{ph}}I_N,
& \partial_{T_{ph}}I_Q^{ph}=\Ome \partial_{T_{ph}}I_N,
\end{align}
and one obtains Eq.~(\ref{aph}). Remarkably, heat current amplification, $\alpha>1$,
is achievable once $|E_\ell|>|\Ome|$ in both the {\em linear} and
nonlinear regimes whenever the inelastic contribution is nonzero. This
conclusion counters present designs in which  a thermal transistor  effect
can only be realized in the nonlinear response regime based on a negative
differential thermal conductance \cite{liwang}. In fact, one can prove
that a thermal transistor effect can materialize in the linear
response regime {\em only when inelastic processes involve more than
  two reservoirs} due to restrictions imposed by the second law of
thermodynamics
[{see Appendix C}]\cite{future}. {The inelastic
  electron hopping mechanism assumed in our three-terminal setup is one of the simplest
 examples which can realize such a nontrivial property: electrons
enter from the source and leave at the drain with the assistance
of phonons from a {\em third} terminal (the substrate).}
Fig.\ref{fig:fig3}(a) demonstrates the thermal transistor
behavior:  $I_Q^L$ changes more significantly than
$I_Q^{ph}$. Specifically, for the adopted temperature difference
$\delta T=0.02 T_0\ll T_0$ and for small $\Delta T = T_{ph}-T_0$ the
system is indeed described by linear response terms. {Although the
coefficient $\alpha$ increases for smaller $\Ome$, the variation of
the heat current $I_Q^L$ as controlled by the temperature of the
phonon terminal can also decrease once the inelastic transport
mechanism weakens [see Fig.~1(b)].}

{In addition to the thermal transistor effect exemplified in Fig.\ref{fig:fig3}(a),
we} also show in Fig.\ref{fig:fig3}(b) that by changing the temperature of the
phonon bath we can substantially modify the $I_e$-$V$ characteristics ---
{This effect is analogous to the field effect transistor, but here it is controlled by the
  phonon temperature instead of the gate voltage at the third
  terminal. This functionality is also different from the
  unconventional transistor effect observed in
  superconductor--normal-metal--superconductor junctions by Saira {\sl
    et al.} where gate-voltage at the third terminal  controlled
  thermal transport between source and drain using the Coulomb
  blockade in the normal-metal region\cite{perkola-trans}.}

\section{Conclusions and discussions}

{We proposed a realistic and relatively simple setup for the realization of thermal,
electrical, and cross rectifiers and transistors by exploiting
phonon-assisted hopping transport in DQD systems in a
three-terminal geometry.} While we used a DQD
system  embedded in a phonon substrate as our platform,
results are applicable to other two-site (two-level)
systems maintained out of equilibrium by multiple biases
and subjected to inelastic transport processes. {Numerical simulations
using realistic parameters demonstrate strong charge,  heat  and
thermoelectric rectifications.} Furthermore, for the first time,
it is observed that a thermal transistor can be realized {\em
 without} negative differential thermal conductance. {In addition,
a transistor effect where the $I_e$-$V$
characteristics is tuned by the phonon temperature was
revealed. These functionalities should enable smart manipulations of
heat and charge transport in nano-systems, core elements in
future information processing technologies. Although calculations
were performed for a specific set of parameters, the uncovered functionalities
should be observed  whenever inelastic transport processes dominate elastic effects.
As discussed in Refs.~\onlinecite{prb1,njp}, this can be
achieved in the high temperature and strong electron-phonon
interaction regime when elastic transport is suppressed e.g. by a barrier, or an energy
gap.
Our study here essentially aimed at the hybridization of two
distinct branches of technology: electronics and
phononics\cite{bli}. Amalgamation of technologies may provide new platforms and
opportunities for high-performance, high-energy-efficiency nanotechnologies.}
Extensions to consider the interplay of spin, charge, and heat
transport in the presence of inelastic effects are of interest for
exploring  spin-caloritronics \cite{spincalorit} devices beyond linear
response. The physics revealed here, i.e., the essential importance of
inelastic transport effects for a linear-response transistor and other
nonlinear device functionalities, is also of importance for 
related research fields.

\section{Acknowledgments}
JHJ acknowledges support from the faculty start-up funding of Soochow
University. MK thanks the hospitality of the Chemical Physics Theory
Group at the Department of Chemistry of the University of Toronto and
the Initiative for the Theoretical Sciences (ITS) - City University of
New York Graduate Center where several interesting discussions took
place during this work. He also gratefully acknowledges support from
the Professional Staff Congress – City University of New York award
No. 68193-0046. DS acknowledges support from an NSERC Discovery Grant
and the Canada Research Chair program. YI acknowledges support from
the Israeli Science Foundation (ISF) and the US-Israel Binational
Science Foundation (BSF).


\begin{appendix}

\section{Derivation of the phonon induced transition rate between DQDs.}

{
The electron-phonon coupling matrix for longitudinal phonons in GaAs
is given by
\be
\zeta_{L,{\vec q}} = e^{i{\vec q}\cdot {\vec
    r}}\sqrt{\frac{\hbar}{2\rho_M v_{sl} qV} } ( e D_{{\vec q},L} - i q
\Xi ) ,\label{eqs1}
\ee
where $V$ is the volume of the system, $\rho_M=5.3\times
10^3$~kg/m$^3$ is the mass density of GaAs, $\Xi=7.0$~eV is the
deformation potential, and $v_{sl}=5.29\times 10^3$~m/s is the velocity of
longitudinal phonons. The piezoelectric potential is
\be
D_{{\vec q},L} = \frac{ 8 \pi e_{14} }{ \kappa} \frac{(3 q_x q_y
  q_z)}{ q^3 } ,
\ee
where $e_{14}=1.41\times 10^9$~V/m is the piezoelectric constant and
$\kappa=12.9$ is the relative dielectric constant. Those material
parameters are adopted from the standard semiconductor handbook,
Ref.~\onlinecite{madelung}. From the above equations, one finds that
\be
|\zeta_{L,{\vec q}}|^2 = \frac{32 \hbar \pi^2 e^2 e_{14}^2}{\kappa^2
  \rho_M v_{sl} q V} \frac{(3q_x q_y q_z)^2}{q^6} + \frac{\hbar
  \Xi^2 q }{ 2 \rho_M v_{sl} V} .
\ee
When considering the coupling of electrons to transverse phonons, only the piezoelectric mechanism
contributes. In this case the two transverse branches of phonons give
the total contribution
\begin{align}
 |\zeta_{T,{\vec q}}|^2 = & \frac{32 \hbar \pi^2 e^2 e_{14}^2}{\kappa^2
  \rho_M v_{st} V q} \nn\\
& \mbox{} \times \left[ \frac{q_x^2 q_y^2 + q_x^2 q_z^2 + q_y^2
  q_z^2}{q^4} - \frac{(3q_x q_y q_z)^2}{q^6} \right] ,
\end{align}
where $v_{st}=2.48\times 10^3$~m/s is the velocity of transverse acoustic phonons.}

{We now consider electronic wavefunctions in the quantum dots, assuming a Gaussian form,
\begin{align}
& \Psi_1({\vec r}) = \frac{1}{(2\pi)^{3/4}l_{qd}^{3/2}}
\exp\left[-\frac{x^2+y^2}{4l_{qd}^2} -
  \frac{(z+d/2)^2}{4l_{qd}^2}\right] , \nn \\
& \Psi_2({\vec r}) = \frac{1}{(2\pi)^{3/4}l_{qd}^{3/2}}
\exp\left[-\frac{x^2+y^2}{4l_{qd}^2} -
  \frac{(z-d/2)^2}{4l_{qd}^2}\right] ,\nn
\end{align}
where $l_{qd}$ is the characteristic length of the wavefunction.
The two wavefunctions are centered at $(0,0,\pm d/2)$ (i.e., the
distance between the two QDs is $d$), respectively.}

\begin{figure}[htbp]
\includegraphics[scale=1]{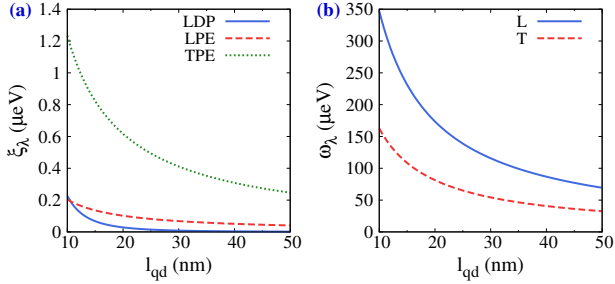}
\caption{(Color online) (a) Electron-phonon coupling strength
  $\xi_\lambda$ vs. quantum dot radius $l_{qd}$ for longitudinal phonon with
  deformation potential coupling mechanism (``LDP''), transverse phonons
  with piezoelectric coupling mechanism (``TPE''), and longitudinal
  phonons with piezoelectric coupling mechanism (``LPE''). Calculated from
  Eqs.~(\ref{a1})-(\ref{a3}) using material parameters of GaAs. (b)
  Characteristic phonon frequency $\ome_\lambda$ as functions of
  quantum dot radius $l_{qd}$ (``L'' and ``T'' stand for longitudinal
  and transverse phonons, respectively).}
\label{fig:sup1}
\end{figure}

{With these elements at hand, we write
down the transition rate between the dots using the Fermi Golden Rule,
\begin{align}
  \gamma_{e-ph} = & \frac{2\pi}{\hbar} \sum_{{\vec q}\lambda} |\zeta_{\lambda,{\vec
      q}}|^2 |\langle \Psi_1 | e^{i{\vec q}\cdot{\vec r}} |
  \Psi_2\rangle |^2 \nn\\
  & \mbox{} \times \delta( \pm \hbar\ome_{\lambda,{\vec q}} - E_r + E_{\ell} ) .
\end{align}
The integral over space gives
\be
|\langle \Psi_1 | e^{i{\vec q}\cdot{\vec r}} |
\Psi_2\rangle |^2  = \exp\left[-\left(q^2 l_{qd}^2 + \frac{d^2}{4l_{qd}^2}\right)\right] .
\ee
The summation over ${\vec q}$ can be converted into integration which
gives
\be
\gamma_{e-ph} =
\sum_{\lambda=LDP,TPE,LPE}\xi_{\lambda}\left(\frac{\Ome}{\ome_{\lambda}}\right)^{n_\lambda}\exp\left[-\left(\frac{\Ome}{\ome_{\lambda}}\right)^2\right]
. \nn
\ee
Here $LDP$ stands for longitudinal phonons and the deformation potential
mechanism, and $TPE$ ($LPE$) stands for transverse (longitudinal)
phonons with the piezoelectric coupling mechanism.
$\ome_{LDP}=\ome_{LPE}=v_{sl}/l_{qd}$, $\ome_{TPE}=v_{st}/l_{qd}$,
$n_{LDP}=3$, $n_{LPE}=n_{TPE}=1$, and
\begin{align}
& \xi_{LDP} = \frac{\Xi^2 }{2\pi \hbar \rho_M
  v_{sl}^2 l_{qd}^3 } \exp\left(-\frac{d^2}{4l_{qd}^2}\right) , \label{a1}\\
& \xi_{LPE} = \frac{96\pi e^2
  e_{14}^2}{35 \hbar \kappa^2 \rho_M v_{sl}^2 l_{qd}}
\exp\left(-\frac{d^2}{4l_{qd}^2}\right) , \label{a2}\\
& \xi_{TPE} = \frac{128\pi e^2
  e_{14}^2}{35 \hbar \kappa^2 \rho_M v_{st}^2
  l_{qd}} \exp\left(-\frac{d^2}{4l_{qd}^2}\right) .
\label{a3}
\end{align}
We calculate $\xi_\lambda$ for GaAs QDs using $d/l_{qd}=4$. Results are
plotted in Fig.~\ref{fig:sup1}. For $l_{qd}\simeq 12$~nm,
$\xi_{TPE}=1$~$\mu$eV and $\ome_{TPE}\simeq 120$~$\mu$eV, approving the parameters, $\xi_0$ and $\ome_0$ (representing $\xi_{TPE}$
and $\ome_{TPE}$ in the main text, since we consider only the piezoelectric
electron-phonon coupling for transverse phonons there), chosen in the
main text.}


\section{Comparison between the Fermi-Golden
 Rule approximation and the Rate equation method}
{
The Fermi Golden Rule approximation is valid when $k_BT \gg
\gamma_{\ell},\gamma_{r}\gg | \Gamma_{\ell\to r} - \Gamma_{r\to
  \ell}|$ \cite{prb1}. Here we will examine situations in which
the condition
$\gamma_{\ell},\gamma_{r}\gg | \Gamma_{\ell\to r} - \Gamma_{r\to
  \ell}|$ is relaxed. 
We assume that the broadenings of the QD levels $\gamma_{\ell},\gamma_{r}$ are
still much smaller than the thermal energy $k_BT$. The transport
currents can then  be calculated through Eq.~(\ref{in-I}). However,
the particle current $I_N$ itself attains a more complex form in terms of
the electrochemical potentials and temperatures. In this regime, the
steady state distributions on the two QDs, $f_\ell$ and $f_r$, can be
calculated by solving the following rate equations in steady state,
\begin{align}
& 0= \frac{d f_\ell}{dt}= - \gamma_\ell [ f_\ell - f_L(E_\ell)] -\gamma_{e-ph}\nn\\
& \quad\quad\quad\quad\quad \times [f_\ell(1-f_r)N_p^- - f_r(1-f_\ell)N_p^+] , \\
& 0= \frac{d f_r}{dt}= - \gamma_r [ f_r - f_R(E_r) ] + \gamma_{e-ph}\nn\\
&\quad\quad\quad\quad\quad \times [f_\ell(1-f_r)N_p^- - f_r(1-f_\ell)N_p^+] .
\end{align}
Once obtaining $f_\ell$ and $f_r$, the particle current can be
calculated via
\be
I_N=\gamma_{e-ph}[f_\ell(1-f_r)N_p^- - f_r(1-f_\ell)N_p^+] .
\ee
When $\gamma_\ell, \gamma_r\gg |I_N|$, the Fermi
Golden Rule approximation is validated, since $f_\ell \simeq
f_L(E_\ell)$ and $f_r \simeq f_R(E_r)$.}

\begin{figure}[htbp]
\includegraphics[scale=0.45]{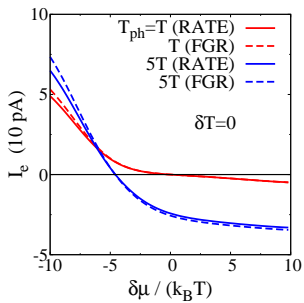}
\caption{ (Color online) $I_e$-$V$ curves for different phonon
  temperatures calculated from the rate equation method (denoted as
  ``RATE'' in the figure) and the
  Fermi Golden Rule (denoted as ``FGR'') approximation. The parameters
  are the same as in  Fig.~\ref{fig:fig3}(b). }
\label{figs2}
\end{figure}

{We plot $I_e$-$V$ characteristics for different
phonon temperatures using the rate equation method and the Fermi Golden
Rule approximation in Fig.~\ref{figs2}. It is noted that
differences are negligible unless we operate in the far-from equilibrium regime,
$T_{ph}\gg T_L, T_R$, or $|\delta\mu|\gg k_BT$. Our results in
Fig.~\ref{figs2} indicate that the nonlinear performance is marginally reduced;
qualitative features remain the same. We speculate that the phenomena
described in this work persist beyond the strict regimes where the rate
equation method or the Fermi Golden Rule approximation are
justifiable. Simulations in the main text
of this paper were calculated using the rate equation method.
}

\section{Restrictions on the thermal transistor effect in the linear-response
  regime from the second-law of thermodynamics}

{
Here we discuss restrictions, directly arising from the second-law of thermodynamics,
on the thermal transistor effect which we materialize in the linear-response regime [see Fig. \ref{fig:fig3}(a)].
Our discussion below is based on general irreversible thermodynamic
arguments, and is holds for any thermodynamic systems in the
linear response regime. Since our analysis here is confined to the thermal transistor effect,
we shall consider pure thermal conduction (i.e., the
electrochemical potential difference is set to zero).}

{For a system with three reservoirs, one can define three heat
  currents (one for each reservoir). However, due to energy conservation,
$I_Q^L+I_Q^R+I_Q^{ph}=0$,
we are left with two  independent heat currents \cite{prb1,jap-j}.
If we choose the heat
  current flowing out of the source and that emerging from the phonon
  terminal as the two independent currents, the thermal transport
  equation for three terminal in the linear response regime can be
  generally written as,
\begin{align}
  \left( \begin{array}{c}
      I_Q^L\\ I_{Q}^{ph}\end{array}\right) =
  \left( \begin{array}{cccc}
      K_L & K_{o} \\
      K_{o} & K_{ph}
    \end{array}\right) \left(\begin{array}{c}
      \frac{T_L - T_R}{T}\\ \frac{T_{ph}- T_R}{T} \end{array}\right) \ .\label{3t-t}
\end{align}
Here $I_Q^L=I_Q^e-I_Q^{ph}/2$ is the heat current flowing out of the
source. $T$ is the equilibrium temperature. $K_L$ and $K_{ph}$ are the
diagonal thermal conductance, while $K_o$ stands for the off-diagonal
thermal conductance. The above equation is derived using Onsager's
theory of linear-response by finding the thermodynamic forces
conjugated to the two heat currents. It can also be obtained by
linearizing the transport equation in the main text, 
Eq.~(\ref{trans-non}), and then reorganizing the transport coefficients
according to the definition of $I_Q^L$. The second-law of
thermodynamics imposes the following restrictions on transport coefficients,
\be
K_L\ge 0, \quad K_{ph} \ge 0, \quad K_L K_{ph} \ge K_o^2 . \label{3t-res}
\ee}

{We now note that in a three-terminal system,
one can generally identify two distinct classes of transport mechanisms:
(i) those involving only two terminals (reservoirs) in each
microscopic process (briefed as ``2T-class''); (ii) those in which
energy is exchanged among three reservoirs in each transport process
(briefed as ``3T-class''). 
}

{For the 2T-class, the thermal transport equations in the
linear response regime are generally given by,
\begin{subequations}
\begin{align}
& I_{L\to R} = K_{LR}(T_L - T_R)/T , \\
& I_{L\to ph} = K_{Lph}(T_L - T_{ph})/T , \\
& I_{R\to ph} = K_{Rph}(T_R - T_{ph})/T ,
\end{align}
\label{2t-trans}
\end{subequations}
where $K_{LR}$, $K_{Lph}$, and $K_{Rph}$ are the thermal conductances
between pairs of reservoirs. The
above equations represent Fourier's law of thermal conduction. The
second law of thermodynamics requires that the thermal conductances are positive,
\be
K_{LR} \ge 0, \quad K_{Lph}\ge 0, \quad K_{Rph} \ge 0 .\label{ge0}
\ee}
{We can organize
   equations (\ref{2t-trans}) into the general form
  (\ref{3t-t}), written in terms of a new set of transport coefficients,
\begin{align}
& K_L^\prime = K_{LR} +  K_{Lph} , \nn\\
&  K_{ph}^\prime = K_{Lph} +
K_{Rph}, \quad K_{o}^\prime = - K_{Lph} .
\end{align}
Based on the restrictions (\ref{ge0}), we find that [on top of Eq. (\ref{3t-res})],
\be
K_{o}^\prime \le 0, \quad |K_{o}^\prime| \le |K_{ph}^\prime| .\label{2t-res}
\ee}

{As discussed in the main text, the heat current amplification factor in the
linear response regime is given by
\be
\alpha \equiv \left |\frac{\partial_{T_{ph}} I_Q^L}{\partial_{T_{ph}}
    I_Q^{ph}} \right| = \left|\frac{K_o}{K_{ph}}\right| .
\ee
From Eq. (\ref{2t-res}) it is now clear that there is no thermal
transistor effect for 2T-class models in the
linear-response regime since they identically result in $\alpha\le 1$.
(Recall, the 2T-class transport mechanisms correspond to a three terminal situation, with transport processes occurring
independently between every pair of terminals. )
In contrast, for 3T-class transport mechanisms,
the restriction (\ref{3t-res}) does {\em not} disallow heat current
amplification, $\alpha>1$. In fact, as discovered in this work, one
can manifest the thermal transistor effect in the linear response regime for the
 phonon-assisted hopping process, which is an example of the 3T-class
transport mechanisms.}

\end{appendix}

\bibliographystyle{apsrev}
\bibliography{qd}

\end{document}